\documentclass[runningheads]{llncs}
\usepackage{graphicx}
\usepackage{verbatim}
\usepackage{array}
\usepackage{multirow}
\usepackage{amsfonts}
\usepackage{stmaryrd}
\usepackage{amsmath}
\usepackage{xcolor}
\usepackage{booktabs,colortbl}
\begin{document}
\title{FetalDiffusion: Pose-Controllable 3D Fetal MRI Synthesis with Conditional Diffusion Model}
\titlerunning{Controllable 3D fetal MRI generation}
%

\author{Molin Zhang\inst{1}
\and
Polina Golland\inst{1,2} \and
P. Ellen Grant\inst{3,4} \and
Elfar Adalsteinsson\inst{1,5}
}

\authorrunning{M. Zhang et al.}
%
\institute{Department of Electrical Engineering and Computer Science, MIT, \\ Cambridge, MA, USA \\\email{molin@mit.edu} \and
Computer Science and Artificial Intelligence Laboratory, MIT, \\
Cambridge, MA, USA \and
Fetal-Neonatal Neuroimaging and Developmental Science Center, \\ Boston Children’s Hospital, Boston, MA, USA \and
Harvard Medical School, Boston, MA, USA \and
Institute for Medical Engineering and Science, MIT, Cambridge, MA, USA}

%
%
\maketitle              
\begin{abstract}
The quality of fetal MRI is significantly affected by unpredictable and substantial fetal motion, leading to the introduction of artifacts even when fast acquisition sequences are employed. The development of 3D real-time fetal pose estimation approaches on volumetric EPI fetal MRI opens up a promising avenue for fetal motion monitoring and prediction. Challenges arise in fetal pose estimation due to limited number of real scanned fetal MR training images, hindering model generalization when the acquired fetal MRI lacks adequate pose.

In this study, we introduce FetalDiffusion, a novel approach utilizing a conditional diffusion model to generate 3D synthetic fetal MRI with controllable pose. Additionally, an auxiliary pose-level loss is adopted to enhance model performance. Our work demonstrates the success of this proposed model by producing high-quality synthetic fetal MRI images with accurate and recognizable fetal poses, comparing favorably with in-vivo real fetal MRI. Furthermore, we show that the integration of synthetic fetal MR images enhances the fetal pose estimation model's performance, particularly when the number of available real scanned data is limited resulting in $15.4\%$ increase in PCK and $50.2\%$ reduced in mean error. All experiments are done on a single 32GB V100 GPU. Our method holds promise for improving real-time tracking models, thereby addressing fetal motion issues more effectively.

\end{abstract}
\section{Introduction}
Fetal MRI faces significant challenges due to the aperiodic, unpredictable, and substantial nature of fetal motion~\cite{jokhi2011magnetic}. Despite efforts to address these issues, such as employing fast acquisition techniques like half-Fourier single-shot rapid acquisition (HASTE)\cite{patel1997half}, and utilizing reconstruction methods that leverage temporal subspace or regularization\cite{zhang2024zero,arefeen2023latent,poddar2015dynamic,biswas2019dynamic}, inter-slice motion can still pose a challenge. Recent advancements in fetal pose estimation approaches~\cite{xu2019fetal,zhang2020enhanced,xu20203d} offer promise for prospective methods aimed at detecting and mitigating fetal motion artifacts through motion tracking. 
The extraction of fetal pose localization facilitates the feasibility of spatially selective excitation for fetal MRI~\cite{zhang2022selective,zhang2023stochastic}, allowing for precise targeting of regions of interest (ROI). 

Challenges arise as the fetal pose estimation model encounters limitations due to a scarcity of real scanned fetal MR training images which may lack sufficient information about fetal pose, necessitating a substantial number of pregnant volunteers for data collection~\cite{xu2019fetal,zhang2020enhanced,xu20203d}. Moreover, the manual labeling of extensive data proves to be a time-consuming and labor-intensive process. It is imperative to generate synthetic high quality of 3D fetal MRI with capability of controllable fetal pose.

In the realm of medical imaging, Generative Adversarial Networks (GANs) play a pivotal role in synthesizing data
~\cite{goodfellow2014generative}. Notably, GANs have demonstrated commendable performance in synthesizing brain imaging data~\cite{han2018gan}. Controllable image synthesis can be achieved through techniques such as conditioning on the discriminator~\cite{mirza2014conditional} or utilizing the CycleGAN framework~\cite{zhu2017unpaired}. These approaches have facilitated diverse applications including contrast-conditioned MRI synthesis~\cite{dar2019image}, edge-aware MRI generation~\cite{yu2019ea}, and CT-MRI translation~\cite{lei2019mri}. In parallel, the Variational Autoencoder (VAE) framework~\cite{kingma2013auto} explores synthetic image generation by learning the compressed data distribution within the latent space. However, GAN-based methods are susceptible to unstable training process and VAE tends to produce blurred images.

Recent breakthroughs in diffusion models and score functions~\cite{ho2020denoising,song2020score} have revolutionized image processing by decomposing the problem into a sequence of forward-backward (diffusion-denoising) operators and demonstrate superior performance compared to GANs~\cite{dhariwal2021diffusion}. 
To address the challenges of high memory usage and extended running times, latent diffusion model (LDM) has been introduced which incorporates an encoder and decoder, enabling the compression of images into compact latent variables
~\cite{rombach2022high}. The successful application of the latent diffusion model in imaging synthesis is evident in various studies, including the generation of 3D brain images~\cite{pinaya2022brain,peng2023generating}, multi-modal MRI synthesis~\cite{jiang2023cola}, image translation~\cite{ozbey2023unsupervised} as well as  person image synthesis~\cite{bhunia2023person} 

This study introduces a pioneering method for achieving controllable 3D fetal MRI based on a specified fetal pose, named FetalDiffusion. Utilizing a conditional diffusion model, our approach operates efficiently on a single 32G V100 GPU. Notably, we are the first to tackle the challenges associated with the scarcity of training data for fetal pose models in conjunction with synthetic fetal MRI models. The key contributions of our work can be summarized as follows:
\begin{enumerate}
    \item We propose a novel 3D diffusion model conditioned on a single 3D mask created by 15 skeleton landmarks and limb areas by cross-attention with high level features.
    \item We add an auxiliary loss using trained fetal pose estimation model with limited data to enforce pose-level constraints.
    \item We demonstrate the efficacy of our method by the production of high-quality synthetic 3D fetal MRI images on both seen and unseen poses. We also show improved pose estimation performance in models trained using the additional data generated by our proposed approach.
\end{enumerate}
Note that LDM is not adopted due to potential loss of fine-grain features in fetal pose landmarks during compression, yielding inferior generative results.

\begin{figure}
    \centering
    \includegraphics[width=0.8\textwidth]{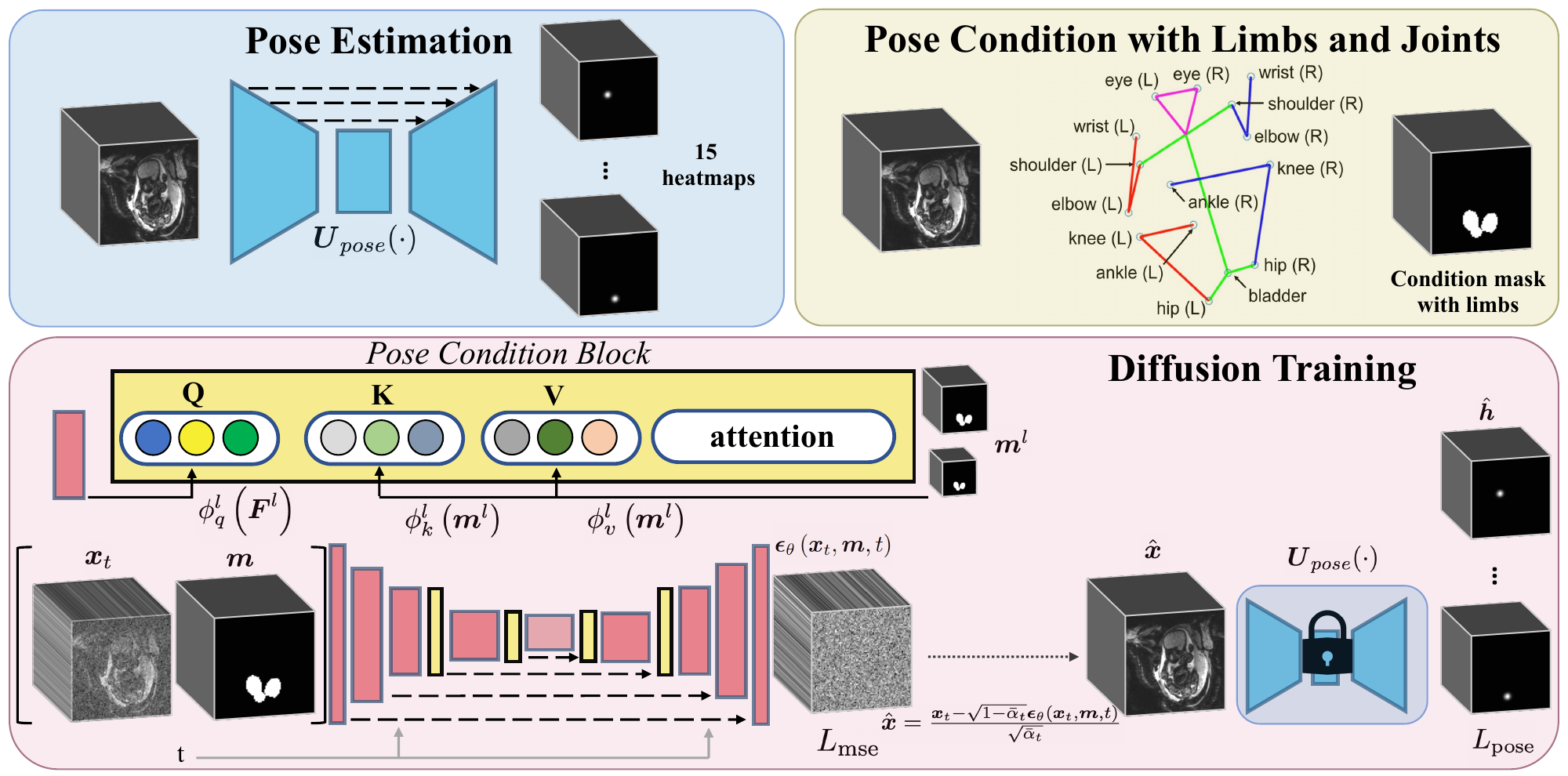}
    \caption{Overall framework. We initially train the pose estimation network using a limited-size dataset. For pose conditioning, we incorporate both landmark spots and limb masks within the condition mask. We follow the same diffusion process as DDPM~\cite{ho2020denoising}. The condition mask and noisy image are concatenated as input for the 3D denoising Unet, featuring four downsampling and upsampling layers (64, 128, 128, 256 channels). Pose Condition Blocks (PCB) are embedded in the last two layers, with the mask downsampled accordingly. For the attention module in PCB, we use 8 heads and the same channel number (128, 256) for each downsampling level at the highest two levels. Using the predicted noise, we directly project back to the image and input it into the trained pose estimation network to create an auxiliary pose loss, enhancing overall performance.}
    \label{fig:fig1}
\end{figure}

\section{Methods}
\subsection{Pose Landmark Preparation and Estimation}
We adopt a previous design choice~\cite{xu2019fetal,zhang2020enhanced}, selecting 15 landmarks, including joints, eyes, and the bladder, as the representation of fetal pose. For fetal pose estimation, generating a confidence heatmap for each landmark using a Gaussian distribution spot proves sufficient. We employ a 3D Unet $\boldsymbol{U}_{pose}(\cdot)$ with a cropped 3D fetal MRI volume as input and 15 predicted heatmap as output. MSE loss is calculated between the predicted heatmaps $\hat{\boldsymbol{h}}$ and groundtruth heatmaps $\boldsymbol{h}$.

In contrast, for the pose-conditional 3D diffusion model, we enhance the representation by incorporating additional feature information related to limbs (2 arms and 2 legs). Given the strong correlation between joints and limbs, this inclusion aims to improve performance of generated synthetic fetal MRI. 
Across different volumes and subjects, variations primarily manifest in the location and orientation of fetal limbs, while the features of the head and body remain consistent. Furthermore, fetal limbs, being thinner than the body and brain, underscore the need for a more detailed conditioning approach. 

Note that using only landmark Gaussian spot heatmaps, without limb information, for the fetal pose estimation network is preferred. This helps avoid directing optimization attention towards a large number of limb voxels as this tends to lead the training towards sub-optimal results. We generate the conditional information limb mask by assigning 1 to the voxel $p$ if it satisfies:

\begin{equation}
\begin{array}{r}
\begin{aligned}
&(p - p^{GT}_k)\cdot a_{km} > 0, \quad
(p - p^{GT}_m)\cdot a_{km} < 0,\\
&\left\|(p - p^{GT}_m)\times a_{km}\right\|_{2}/\left\|a_{km}\right\|_{2} \leq r
\end{aligned}
\end{array}
\end{equation}
where $p^{GT}_k$ and $ p^{GT}_m$ are two landmark locations linked by a limb and $a_{k,m} = p^{GT}_k- p^{GT}_m$. In this work we use $r = 6$.

\subsection{Pose-Conditional 3D Diffusion Model}

\subsubsection{Overall 3D Diffusion Framework}
Fig.~\ref{fig:fig1} shows the overview of the proposed generative model of FetalDiffusion. Our goal is to train the diffusion model to learn the data distribution of 3D fetal MRI $q\left(\boldsymbol{x}\right)$ given a condition of fetal pose mask $\boldsymbol{m}$.
The denoising network $\boldsymbol{\epsilon}_\theta$ is a Unet-based architecture where we condition the pose information at the input by concatenating the noisy input $\boldsymbol{x}_t$ and $\boldsymbol{m}$ at diffusion step $t$. 

The generative modeling scheme of FetalDiffusion is based on the Denoising diffusion probabilistic model (DDPM)~\cite{ho2020denoising}. The forward diffusion process is modeled as a Markov chain with the following conditional distribution:
\begin{equation}
    q\left(\boldsymbol{x}_t|\boldsymbol{x}_{t-1}\right)=\mathcal{N}\left(\boldsymbol{x}_t ; \sqrt{1-\beta_t} \boldsymbol{x}_{t-1}, \beta_t \mathbf{I}\right)
\end{equation}
where $t \sim[1, T]$ and $\beta_1, \beta_2, \ldots, \beta_T$ is a series of scaled-linear scheduled variance with $\beta_t \in(0,1)$. We define $\alpha_t=1-\beta_t$ and $\bar{\alpha}_t=\prod_{i=1}^t \alpha_i$. For arbitrary diffusion step $t$, it can be derived that $\boldsymbol{x}_t=\sqrt{\bar{\alpha}_t} \boldsymbol{x}_0+\sqrt{1-\bar{\alpha}_t}\boldsymbol{\epsilon}$ where $\boldsymbol{\epsilon} \sim \mathcal{N}(0, \mathbf{I})$. For the backward denoising process, the posterior can be approximated by a learned deep network. Based on the derivation in~\cite{ho2020denoising}, the training loss for the diffusion model can be depicted in respect to the prediction of the noise $\boldsymbol{\epsilon}$ as follows:
\begin{equation}
    L_{\mathrm{mse}}=\mathbb{E}_{t \sim[1, T], \boldsymbol{x}_0 \sim q\left(\boldsymbol{x}_{\mathbf{0}}\right), \epsilon}\left\|\boldsymbol{\epsilon}-\boldsymbol{\epsilon}_\theta\left(\boldsymbol{x}_t, \boldsymbol{m}, t\right)\right\|^2
\end{equation}
where $\boldsymbol{\epsilon}_\theta\left(\boldsymbol{x}_t, \boldsymbol{m}, t\right)$ is the 3D Unet denoising model taking noisy image $\boldsymbol{x}_t$, conditional information $\boldsymbol{m}$ and diffusion step $t$ as input.

To integrate conditional pose information into the diffusion process, we introduce Pose Condition Blocks (PCB) utilizing a cross-attention mechanism. These blocks are embedded into layers with varying scales or resolutions. Due to significant GPU memory requirements for 3D volumes, we opt to incorporate PCB by downsampling the condition mask only into the highest two layers at a coarser resolution. This choice is rational as it mitigates large GPU memory and computation demands at lower layers, while the robustness of limb-based features are preserved at higher levels. 

\begin{figure}
    \centering
    \includegraphics[width=\textwidth]{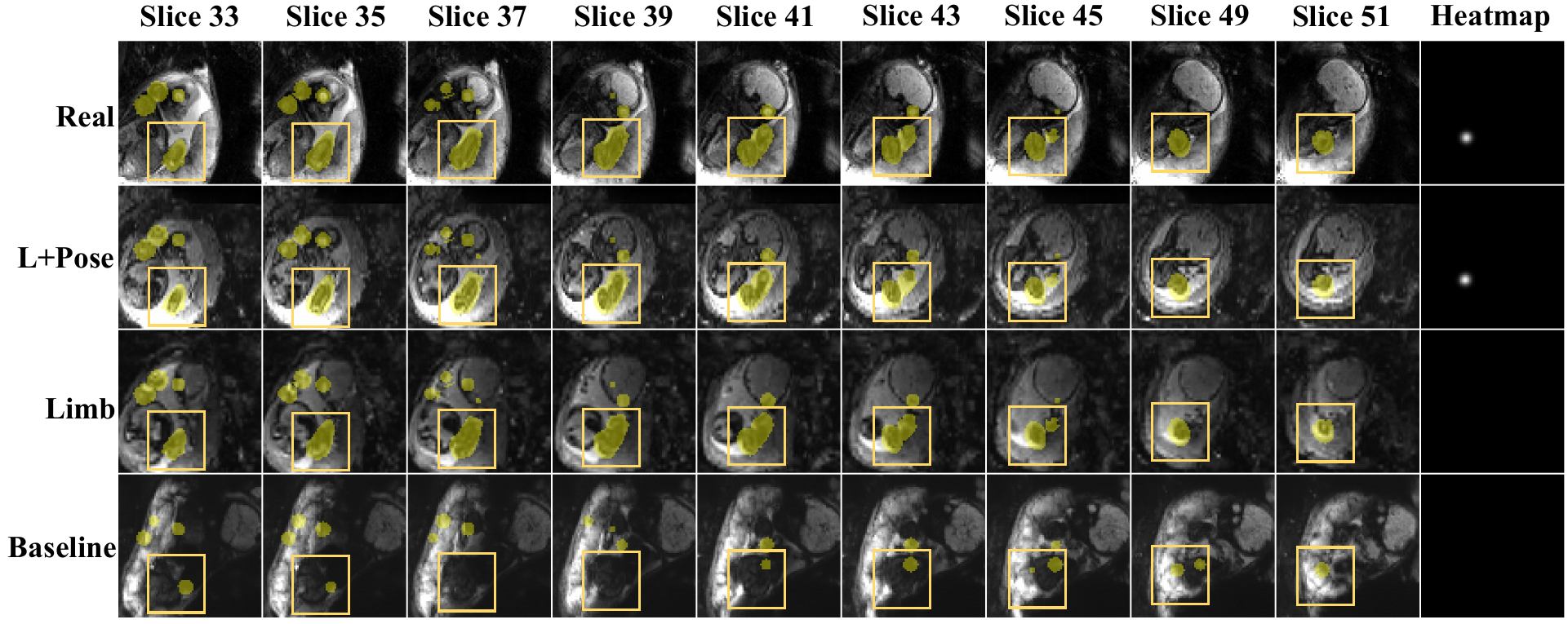}
    \caption{Illustration of synthetic data given a pose from the training dataset is shown below. Rows represent images from real scan data, our proposed method, limb mask without pose loss, and the baseline method using a landmark spot mask. Columns are slices at the z-direction. The last column displays pose estimation results on slice 51 (where the target landmark, shoulder, is located) from the trained network. Our proposed method generates high-fidelity data with correct limb and landmark positioning in the condition mask (light yellow mask) and is detectable by the trained pose estimation network. While the limb mask condition can generate limbs under the condition mask, the right 'arm' is not attached to the fetal body, making it undetectable by the pose estimation network, as indicated by the yellow box. The baseline fails to follow the condition information and does not generate convincing images.}
    \label{fig:fig2}
\end{figure}

\subsubsection{Pose Condition Blocks (PCB)}
To guide the diffusion model with conditional pose information, we integrate cross-attention-based PCB into both the encoder and decoder parts of the Unet architecture across various scale levels of feature layer. The keys $\boldsymbol{K}$ and values $\boldsymbol{V}$ are derived from $\boldsymbol{m}^l$ that is downsampled $\boldsymbol{m}$ at the corresponding resolution of the embedded layer. $l$ denotes the level of the layer. The queries Q are obtained from the features of the layers $\boldsymbol{F}^l$. The cross-attention formula is shown as follows:
\begin{equation}
\begin{array}{r}
\boldsymbol{Q}=\phi_q^l\left(\boldsymbol{F}^l\right), \boldsymbol{K}=\phi_k^l\left(\boldsymbol{m}^l\right),\boldsymbol{V}=\phi_v^l\left(\boldsymbol{m}^l\right), \boldsymbol{F}_{\boldsymbol{o}}^{\boldsymbol{l}}=\boldsymbol{W}^{\boldsymbol{l}} \operatorname{softmax}\left(\frac{\boldsymbol{Q} \boldsymbol{K}^T}{\sqrt{C}}\right) \boldsymbol{V}+\boldsymbol{F}_h^l
\end{array}
\end{equation}
where $\phi_q^l$, $\phi_k^l$, $\phi_v^l$ denote linear layers used to standardize the attention dimension and $\boldsymbol{W}^{\boldsymbol{l}}$ represents a trainable weight.

\begin{table}[h]
\caption{PCK performance (error $<1.2$cm) of different models on 100 randomly selected training samples.}
\label{tab1}
\begin{center}
\begin{tabular}{c|c|ccccccccc}
\hline
metric &method & eye &  shoulder& elbow  & wrist & bladder & hip & knee & ankle & all\\
\hline
\multirow{4}{*}{\shortstack{PCK\\(\%)$\uparrow$}}
        &Real  &\textit{100.0} &   \textit{100.0}   &  \textit{99.0}          &   \textit{100.0}        &\textit{100.0}        &\textit{93.0}   &\textit{99.0}     &\textit{96.0}     &\textit{98.3}  \\
&Baseline  &5.0  &4.5 &3.0  &5.0  &12.0 &2.0    &6.0    &3.5      &4.7  \\
&Limb    &96.0     &93.5   &63.0   &61.0    &93.0   &93.0     &65.0     &64.0     & 77.6 \\
&L+Pose  &\textbf{97.0}     &\textbf{99.5}      &\textbf{92.0}  &\textbf{87.5}    &\textbf{100.0}          &\textbf{99.5}       &\textbf{90.0}     &\textbf{79.5}     &\textbf{92.7}\\
\hline
\end{tabular}
\end{center}
\end{table}

\begin{figure}
    \centering
    \includegraphics[width=\textwidth]{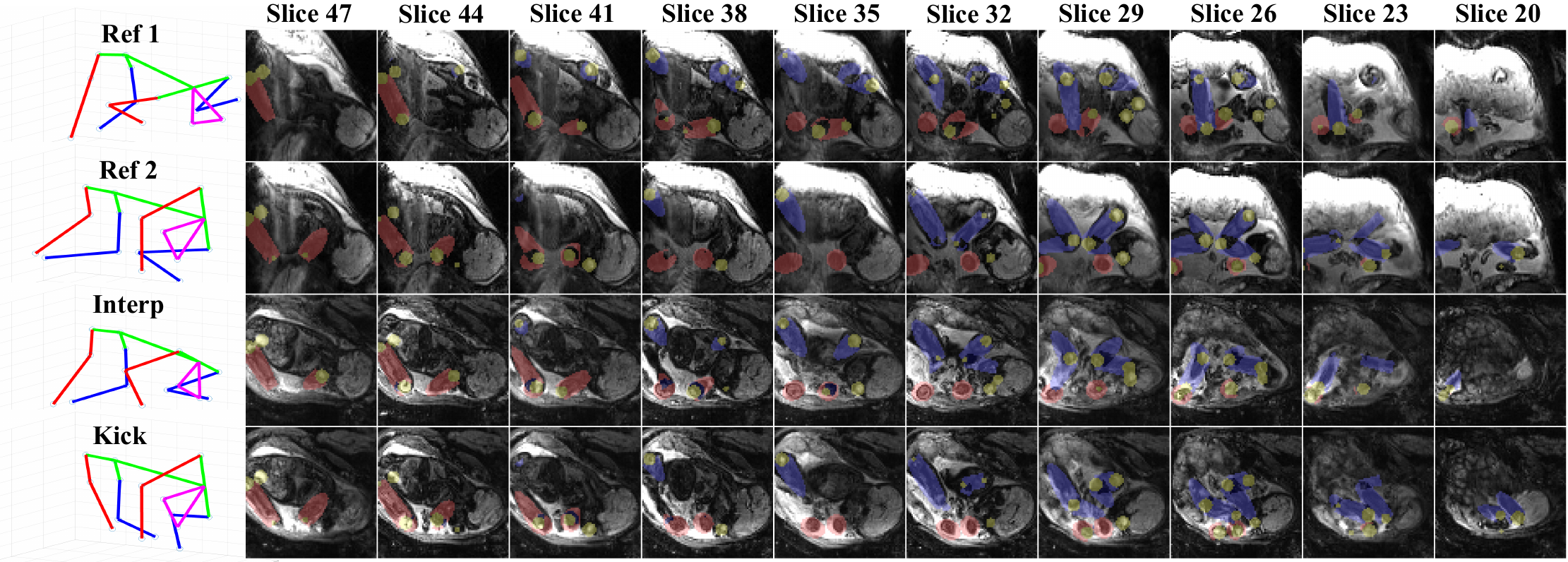}
    \caption{Illustration of synthetic data from our proposed method with an unseen pose is presented in the first column. Rows 1 and 2 depict two reference poses from the training dataset and their corresponding real scanned data. The third row displays an artificially created pose by center interpolating pose ref 1 and ref 2. The fourth row showcases a manually created pose from ref 2, simulating a kicking action in legs and elbows. The color mask represents the condition mask for the diffusion model, with red for left limbs, blue for right limbs, and yellow for landmarks. Our proposed method generates high-fidelity and controllable images for these unseen poses.}
    \label{fig:fig3}
\end{figure}

\subsection{Auxiliary Pose-Level Loss}
The primary objective of this study is to produce high-quality synthetic fetal MRI conditioned on any given fetal pose, particularly beneficial for enhancing the performance of the pose estimation network $\boldsymbol{U}{pose}(\cdot)$ in scenarios with limited training data. While the synthetic data may exhibit high quality, its utility for pose estimation is not guaranteed. To address this, we propose an innovative auxiliary pose-level loss. In this approach, we input the generated image into $\boldsymbol{U}{pose}(\cdot)$ and calculate the loss by comparing the output landmark heatmaps with the ground truth heatmaps. 
Instead of using the complete sampled image from the reverse diffusion which takes 3 minute for one sample, we use $\hat{\boldsymbol{x}}= \frac{\boldsymbol{x}_t - \sqrt{1-\bar{\alpha}_t}\boldsymbol{\epsilon}_\theta\left(\boldsymbol{x}_t, \boldsymbol{m}, t\right)}{\sqrt{\bar{\alpha}_t} }$

\begin{equation}
    L_{\mathrm{pose}}=\mathbb{E}_{t \sim[1, T], \boldsymbol{x}_0 \sim q\left(\boldsymbol{x}_{\mathbf{0}}\right), \epsilon}\left\|\boldsymbol{U}_{pose}\left(\hat{\boldsymbol{x}}\right) - \boldsymbol{h}\right\|^2    
\end{equation}
The total loss is $L = L_{\mathrm{mse}} + \lambda L_{\mathrm{pose}}$, where $\lambda$ is the coefficient and we use 0.1.
\section{Experiments and Results}
\subsection{Dataset}
The dataset consists 56 3D BOLD MRI (15,148 volumes) acquired on a 3T Skyra scanner (Siemens  Healthcare,Erlangen,  Germany) with multislice, single-shot, gradient echo EPI sequence. The in-plane resolution is $3\times3mm^2$ and slice thickness is $3mm$. 
The gestational age range of the 56 fetuses ranged from 25 to 35 weeks. 
TR=$5-8$s, TE=$32-38$ms, FA=90$^{\circ}$. 28 fetuses, 7,664 volumes were used for pose estimation training in the case of sufficient training data. A subset of $39\%$ of the training dataset (12 fetuses, 3,014 volumes) are used to simulate limited pose estimation training data scenario. 14 fetuses, 3,402 volumes were used for validation. 14 fetuses, 4,082 volumes were used for testing. Random rotation and flip are used as data augmentation methods. The volume is center cropped into $80\times80\times80$ to save memory.

\subsection{Experiments setup}
Our generative diffusion model is trained on a small dataset. We evaluate the model on three aspects: 

\textbf{1.} Training dataset evaluation, considering data visualization, condition accuracy, and pose estimation using Percentage of Correct Keypoint (PCK). 

\textbf{2.} Generalization on unseen and artificially posed fetal data. 

\textbf{3.} Generating additional synthetic data to supplement the limited training dataset, re-training the pose estimation network, and evaluating results using PCK and mean error.

We perform two ablation studies: 1. Baseline diffusion model, denoted as '\textbf{\textit{Baseline}}', utilizing landmark Gaussian spot heatmaps as condition information. 2. Diffusion model incorporating both landmark Gaussian spots and limb masks, denoted as '\textbf{\textit{Limb}}'. Our proposed method combines these condition information aspects and introduces an additional auxiliary loss using the trained pose estimation network during training, denoted as '\textbf{\textit{L+Pose}}'.

The pose estimation network undergoes training for 60 epochs (4 hours) with initial learning rate (lr) at $5e-5$ and cosine decay, and the model with the best validation loss is chosen. The architecture is the same as in~\cite{xu2019fetal}. The diffusion model uses a 3D Unet with proposed PCB blocks. We use 1000 diffusion steps. The model is trained until the loss no longer decreases with lr at $1e-5$. The training process takes about 5 days and the sampling for one data takes 3 mins. The diffusion model is implemented based on MONAI~\cite{cardoso2022monai} using the architecture of 'DiffusionModelUNet'. All experiments are conducted on a single 32G V100 GPU card. Batch size for diffusion model is 4.

\begin{table}[h]
\caption{PCK performance (error $<9$ mm) and mean error on test dataset.}
\label{tab2}
\begin{center}
\begin{tabular}{c|c|ccccccccc}
\hline
metric &method & eye &  shoulder& elbow  & wrist & bladder & hip & knee & ankle & all\\
\hline
\multirow{5}{*}{\shortstack{PCK\\(\%)$\uparrow$}}
&Full  &\textit{98.7} &   \textit{99.8}   &  \textit{95.9}          &   \textit{82.7}        &\textit{98.0}        &\textit{95.0}   &\textit{97.4}     &\textit{82.5}     &\textit{93.4}  \\
&Limited              &96.1              &97.8          &76.2  &58.3           &90.1 &59.2    &78.3    &32.8      &72.4  \\
&Limb    &98.3     &98.8               &80.8   &63.5    &94.5          &91.6       &91.8     &37.4     & 81.2 \\
&L+Aux  &\textbf{98.5}     &\textbf{99.8}      &\textbf{92.6}  &\textbf{75.6}    &\textbf{98.8}          &\textbf{92.9}       &\textbf{96.0}     &\textbf{53.9}     &\textbf{87.8}\\
\hline
\multirow{5}{*}{\shortstack{Mean\\(mm)$\downarrow$}}
&Full  &\textit{2.49} &   \textit{1.86}   &  \textit{4.13}          &   \textit{10.70}        &\textit{2.41}        &\textit{3.18}   &\textit{3.80}     &\textit{11.11}     &\textit{5.13}  \\
&Limited              &3.63              &4.15            &17.25  &27.03           &11.17 &21.25    &12.71    &32.82     &16.59  \\
&Limb    &2.86     &2.95               &13.30   &23.91    &7.26          &5.21       &6.92     &27.65     & 11.53 \\
&L+Aux  &\textbf{2.50}     &\textbf{1.99}      &\textbf{6.63}  &\textbf{15.70}    &\textbf{3.89}          &\textbf{5.09}       &\textbf{5.14}     &\textbf{22.99}     &\textbf{8.26}\\
\hline
\end{tabular}
\end{center}
\end{table}

\section{Results and Discussions}
Fig.~\ref{fig:fig2} illustrates synthetic data generated by our proposed method (L + Pose) given a fetal pose from the training dataset. This method produces high-quality data with reasonable pose and limbs accurately positioned within the condition mask. The pose can be detected by the trained pose estimation network effectively, as evident from the similar heat spot (left shoulder) in the heatmap. 

In contrast, the Baseline method fails to adhere to the condition, highlighting the inadequacy of a simple landmark Gaussian spot mask as the condition information. This simplistic mask fails to capture the relative relationships between different body parts, leading to inaccuracies such as generating the fetal knee at the location of the shoulder. 

The Limb mask condition generates high-fidelity limbs but faces detachment issues without auxiliary pose loss where the arm is not connected to the body. The introduction of pose loss aids in exploiting the relationships between body parts, rectifying the arm detachment issue, as shown in the yellow box.

Table~\ref{tab1} further demonstrates that our proposed method achieves high-fidelity data, aligning well with the trained fetal pose estimation network.

Fig.~\ref{fig:fig3} depicts the synthetic data generated by our proposed method under unseen fetal pose conditions. We achieved this by center interpolating reference poses 1 and 2 and adjusting the ankles and elbows of the fetus from reference 2 to simulate a kicking action. These poses do not appear in the training dataset and are realistic. Our generated data accurately adheres to the pose conditions, with limbs and landmarks precisely positioned within the color masks. As a results of consistent contrast and fetal body structure, the diffusion model effectively learns the data distribution with emphasis on pose using the limited size training dataset. More successful illustrations and failure synthetic data with unrealistic pose are shown in supplementary materials.

Table~\ref{tab2} presents the pose estimation results based on different training dataset sizes: full (7,664), limited (3,014), limited with an additional 856 volumes (3,870) augmented from the limb method, and our proposed method. The augmented data is derived from poses in the training dataset and pose diverse as much as possible. The baseline method with low-accuracy synthetic data, is excluded from the analysis. Our proposed method demonstrates substantial improvements compared to the limited dataset in both PCK (15.4\%) and mean error (50.2\%) metrics. Despite utilizing only 28\% additional data, the poses in this additional synthetic set exhibit high pose variation compared to real scans where the fetus is not in motion for most of the time.





\section{Conclusion}
 In this work we propose FetalDiffusion to generate 3D synthetic fetal MRI with controllable fetal pose through a conditional diffusion model. Our model showcases success in producing high-quality, controllable images and exhibits improvements in pose estimation with a limited amount of training data.

%
%
\bibliographystyle{splncs04}
\bibliography{reference}
%




\end{document}